\documentstyle[psfig,12pt]{article}

\def\etal{{\it et al.}}
\def\ie{{\it i.e.}}
\def\bold#1{\setbox0=\hbox{$#1$}
     \kern-.025em\copy0\kern-\wd0
     \kern.05em\copy0\kern-\wd0
     \kern-.025em\raise.0433em\box0 }
\def\half{{\textstyle{1 \over 2}}}

\def\quarter{{\textstyle{1 \over 4}}}
\def\fourth{{\textstyle{1 \over 4}}}

\def\eighth{{\textstyle{1 \over 8}}}

\def\beq{\begin{equation}}
\def\eeq{\end{equation}}
\def\bea{\begin{eqnarray}}
\def\eea{\end{eqnarray}}
\def\bq{\begin{quote}}
\def\eq{\end{quote}}

\def\bq{\begin{quote}}
\def\eq{\end{quote}}

\def\bq{\begin{quote}}
\def\eq{\end{quote}}

\def \lsim{\mathrel{\vcenter
     {\hbox{$<$}\nointerlineskip\hbox{$\sim$}}}}
\def \gsim{\mathrel{\vcenter
     {\hbox{$>$}\nointerlineskip\hbox{$\sim$}}}}

\def\gappeq{\mathrel{\rlap {\raise.5ex\hbox{$>$}}
{\lower.5ex\hbox{$\sim$}}}}

\def\lappeq{\mathrel{\rlap{\raise.5ex\hbox{$<$}}
{\lower.5ex\hbox{$\sim$}}}}

\def\bbz{fa Z \kern-8.9pt Z}

\begin{document}

\baselineskip 18pt
\newcommand{\sheptitle}
{Anomaly Mediated Supersymmetry Breaking without $\boldmath{R}$--Parity}

\newcommand{\shepauthor} {F. de Campos${}^1$, M. A. D\'{\i}az${}^2$,
O. J. P. \'Eboli${}^3$, M. B. Magro${}^3$, and P. G. Mercadante$^3$}

\newcommand{\shepaddress}
{
${}^1$ Departamento de F\'{\i}sica e Qu\'{\i}mica, Universidade Estadual 
       Paulista,\\
       Av. Dr. Ariberto Pereira da Cunha 333, Guaratinguet\'a, SP, Brazil\\
${}^2$ Departamento de F\'{\i}sica, Universidad Cat\'olica de Chile, \\
       Av. Vicu\~na Mackenna 4860, Santiago, Chile\\
${}^3$ Instituto de F\'{\i}sica, Universidade de S\~ao Paulo, CP 66.318, \\
       05389--970, S\~ao Paulo, SP, Brazil 
}

\newcommand{\shepabstract} { We analyze the low energy features of a
  supersymmetric standard model where the anomaly--induced contributions to
  the soft parameters are dominant in a scenario with bilinear $R$--parity
  violation. This class of models leads to mixings between the standard model
  particles and supersymmetric ones which change the low energy phenomenology
  and searches for supersymmetry. In addition, $R$--parity violation
  interactions give rise to small neutrino masses which we show to be
  consistent with the present observations. }

\begin{titlepage}
\begin{flushright}
UCCHEP/17--01\\
IFUSP--1526
\end{flushright}
\begin{center}
{\large{\bf \sheptitle}}
\bigskip \\ \shepauthor \\ \mbox{} \\ {\it \shepaddress} \\ \vspace{.5in}
{\bf Abstract} \bigskip \end{center} \setcounter{page}{0}
\shepabstract
\end{titlepage}

\section{Introduction}

Supersymmetry (SUSY) is a promising candidate for physics beyond the Standard
Model (SM) and there is a large ongoing search for supersymmetric partners of
the SM particles.  However, no positive signal has been observed so far.
Therefore, if supersymmetry is a symmetry of nature, it is an experimental
fact that it must be broken. The two best known classes of models for
supersymmetry breaking are gravity--mediated \cite{GravMed} and gauge--mediated
\cite{GaugMed} SUSY breaking. In gravity--mediated models, SUSY is assumed to
be broken in a hidden sector by fields which interact with the visible
particles only via gravitational interactions and not via gauge or Yukawa
interactions. In gauge--mediated models, on the contrary, SUSY is broken in a
hidden sector and transmitted to the visible sector via SM gauge interactions
of messenger particles.

There is a third scenario, called anomaly--mediated SUSY breaking
\cite{AnomMed}, which is based on the observation that the super--Weyl anomaly
gives rise to loop contribution to sparticle masses. The anomaly contributions
are always present and in some cases they can dominate; this is the anomaly
mediated supersymmetry breaking (AMSB) scenario. In this way, the gaugino
masses are proportional to their corresponding gauge group $\beta$--functions
with the lightest SUSY particle being mainly wino.  Analogously, the scalar
masses and trilinear couplings are functions of gauge and Yukawa
$\beta$--functions. Without further contributions the slepton squared masses
turn out to be negative. This tachyonic spectrum is usually cured by adding an
universal non--anomaly mediated contribution $m_0^2>0$ to every scalar mass
\cite{GGW}.

So far, most of the work on AMSB has been done assuming $R$--Parity ($R_P$)
conservation \cite{others,phenom,fengmoroi}; see \cite{AD} for an exception.
$R$--Parity violation \cite{Allanachetal} has received quite some attention
lately motivated by the SuperKamiokande collaboration results on neutrino
oscillations \cite{NeuAnom}, which indicate neutrinos have mass
\cite{Fukuda:1998mi}. One way of introducing mass to the neutrinos is via
Bilinear $R$--Parity Violation (BRpV) \cite{BRpVrecent}, which is a simple and
predictive model for the neutrino masses and mixing angles
\cite{Hempfling,numassBRpV}. In this work, we study the phenomenology of an
anomaly mediated SUSY breaking model which includes Bilinear $R$--Parity
Violation (AMSB--BRpV), stressing its differences to the $R$--Parity conserving
case.

In BRpV--MSSM \cite{BRpVothers}, bilinear $R$--parity and lepton number
violating terms are introduced explicitly in the superpotential. These terms
induce vacuum expectation values (vev's) $v_i$ for the sneutrinos, and
neutrino masses through mixing with neutralinos. At tree level, only one
neutrino acquires a mass \cite{santamaria}, which is proportional to the
sneutrino vev in a basis where the bilinear $R$--Parity violating terms are
removed from the superpotential. At one--loop, three neutrinos get a non--zero
mass, producing a hierarchical neutrino mass spectrum \cite{SugraBRpV}. It has
been shown that the atmospheric mass scale, given by the heaviest neutrino
mass, is determined by tree level physics and that the solar mass scale, given
by the second heaviest neutrino mass, is determined by one--loop corrections
\cite{numassBRpV}.

In our model, the presence of $R_P$ violating interactions gives rise to
neutrino masses which we show to be consistent with the present observations.
Moreover, the low--energy phenomenology is quite distinct of the conserving
$R$--Parity AMSB scenario. For instance, the lightest supersymmetric particle
(LSP) is unstable, which allows regions of the parameter space where the stau
or the tau--sneutrino is the LSP. In our scenario, decays can proceed via the
mixing between the standard model particles and supersymmetric ones. As an
example, the mixing between the lightest neutralino $\tilde{\chi}^0_1$
(chargino $\tilde{\chi}^\pm_1$) and $\nu_{\tau}$ ($\tau^\pm$) allows the
following decays
\begin{eqnarray*}
   \tilde{\chi}^0_1 &&\to \nu_\tau Z^* \; , 
\\
   \tilde{\chi}^0_1 &&\to \tau^\pm W^{\mp*} \; ,
\\
   \tilde{\chi}^\pm_1 &&\to \tau^\pm Z^* \; ,
\\
   \tilde{\chi}^\pm_1 &&\to \nu_\tau  W^{\pm*}\; .
\end{eqnarray*}

Another effect of the mixing between the standard model and supersymmetric
particles is a sizeable change in the mass of the supersymmetric particles.
For instance, the mixing between scalar taus and the charged Higgs can lead to
an increase in the splitting between the two scalar tau mass eigenstates by a
factor that can be as large as 10 with respect to the $R_P$ conserving case.

This paper is organized as follows. We define in Sec.\ \ref{model} our anomaly
mediated SUSY breaking model which includes Bilinear $R$--Parity Violation,
stating explicitly our working hypotheses. This Section also contains an
overall view of the supersymmetric spectrum in our model. We study the
properties of the CP--odd, CP--even, and charged scalar particles in Sections
\ref{cpodd}, \ref{cpeven}, and \ref{charged} respectively, concentrating on
the mixing angles that arise from the introduction of the $R$--Parity violating
terms. Section \ref{neutrino} contains the analysis that shows that our model
can generate neutrino masses in agreement with the present knowledge.  In
Sec.\ \ref{discussion} we provide a discussion of the general phenomenological
aspects of our model while in Sec. \ref{conclusion} we draw our conclusions.

\section{The AMSB--BRpV model}
\label{model}

Our model, besides the usual $R_P$ conserving Yukawa terms in the
superpotential, includes the following bilinear terms
\begin{equation}
W_{bilinear}=-\varepsilon_{ab}\left(\mu\widehat H_d^a\widehat H_u^b+
\epsilon_i\widehat L_i^a\widehat H_u^b\right)\;,
\end{equation}
where the second one violates $R_P$ and we take $|\epsilon_i|\ll|\mu|$.
Analogously, the relevant soft bilinear terms are
\begin{eqnarray}
V_{soft}&=&m_{H_u}^2H_u^{a*}H_u^a+m_{H_d}^2H_d^{a*}H_d^a+
M_{L_i}^2\widetilde L_i^{a*}\widetilde L_i^a -\nonumber\\
&&-\varepsilon_{ab}\left(
B\mu H_d^aH_u^b+B_i\epsilon_i\widetilde L_i^aH_u^b\right)\;,
\end{eqnarray}
where the terms proportional to $B_i$ are the ones that violates $R_P$.  The
explicit $R_P$ violating terms induce vacuum expectation values $v_i$,
$i=1,2,3$ for the sneutrinos, in addition to the two Higgs doublets vev's
$v_u$ and $v_d$. In phenomenological studies where the details of the neutrino
sector are not relevant, it has been proven very useful to work in the
approximation where $R_P$ and lepton number are violated in only one
generation \cite{BRpB_tau}. In these cases, a determination of the mass scale
of the atmospheric neutrino anomaly within a factor of two is usually enough,
and that can be achieved in the approximation where $R_P$ is violated only in
the third generation.

In this work we assume that $R_P$ violation takes place only in the third
generation, and consequently the parameter space of our model is
\begin{equation}
m_0\,,\, m_{3/2}\,,\, \tan\beta\,,\, {\mathrm{sign}}(\mu)\,,\, 
\epsilon_3\,,\, {\mathrm{and}}\,\, m_{\nu_{\tau}}\,,
\end{equation}
where $m_{3/2}$ is the gravitino mass and $m_0^2$ is the non--anomaly mediated
contribution to the soft masses needed to avoid the appearance of tachyons.
We characterize the BRpV sector by the $\epsilon_3$ term in the superpotential
and the tau--neutrino mass $m_{\nu_{\tau}}$ since it is convenient to trade
$v_3$ by $m_{\nu_{\tau}}$.

In AMSB models, the soft terms are fixed in a non--universal way at the
unification scale which we assumed to be $ M_{GUT} = 2.4 \times 10^{16}$ GeV;
see Appendix~\ref{appa} for details. We considered the running of the masses
and couplings to the electroweak scale, assumed to be the top mass, using the
one--loop renormalization group equations (RGE) that are presented in
Appendix~\ref{appb}. In the evaluation of the gaugino masses, we included the
next--to--leading order (NLO) corrections coming from $\alpha_s$, the
two--loop top Yukawa contributions to the beta--functions, and threshold
corrections enhanced by large logarithms; for details see \cite{GGW}. The NLO
corrections are especially important for $M_2$, leading to a change in the
wino mass by more than 20\%.

One of the virtues of AMSB models is that the $SU(2) \otimes U(1)$ symmetry is
broken radiatively by the running of the RGE from the GUT scale to the weak
one. This feature is preserved by our model since the one--loop RGE are not
affected by the bilinear $R_P$ violating interactions; see
Appendix~\ref{appb}.  In our model, the electroweak symmetry is broken by the
vacuum expectation values of the two Higgs doublets $H_d$ and $H_u$, and the
neutral component of the third left slepton doublet $\widetilde L_3$. 
We denote these fields as
\begin{eqnarray} 
&&H_d={{{1\over{\sqrt{2}}}[\chi^0_d+v_d+i\varphi^0_d]}\choose{ 
H^-_d}}\,,\qquad 
H_u={{H^+_u}\choose{{1\over{\sqrt{2}}}[\chi^0_u+v_u+ 
i\varphi^0_u]}}\,, 
\nonumber \\ 
&&\qquad\qquad\qquad\widetilde L_3={{{1\over{\sqrt{2}}} 
[\tilde\nu^R_{\tau}+v_3+i\tilde\nu^{i0}_{\tau}]}\choose{\tilde\tau^-}}\,. 
\label{eq:shiftdoub} 
\end{eqnarray}

The above vev's $v_i$ can be obtained through the minimization conditions, or
tadpole equations, which in the AMSB--BRpV model are
\begin{eqnarray}
t_d^0&=&(m_{H_d}^2+\mu^2)v_d-B\mu v_u-\mu\epsilon_3v_3+
\eighth(g^2+g'^2)v_d(v_d^2-v_u^2+v_3^2)\,,
\nonumber \\
t_u^0&=&(m_{H_u}^2+\mu^2+\epsilon_3^2)v_u-B\mu v_d+B_3\epsilon_3v_3-
\eighth(g^2+g'^2)v_u(v_d^2-v_u^2+v_3^2)\,, \nonumber \\
t_3^0&=&(m_{L_3}^2+\epsilon_3^2)v_3-\mu\epsilon_3v_d+B_3\epsilon_3v_u+
\eighth(g^2+g'^2)v_3(v_d^2-v_u^2+v_3^2)\,, \label{eq:tadpoles} 
\end{eqnarray}
at tree level.  At the minimum we must impose $t^0_d=t^0_u=t^0_3=0$. In
practice, the input parameters are the soft masses $m_{H_d}$, $m_{H_u}$, and
$m_{L_3}$, the vev's $v_u$, $v_d$, and $v_3$ (obtained from $m_Z$,
$\tan\beta$, and $m_{\nu_{\tau}}$), and $\epsilon_3$. We then use the tadpole
equations to determine $B$, $B_3$, and $|\mu|$.

One--loop corrections to the tadpole equations change the value of $|\mu|$ by
${\cal O}(20\%)$, therefore, we also included the one--loop corrections due to
third generation of quarks and squarks \cite{SugraBRpV}:
\begin{equation}
t_i=t_i^0+\tilde T_i(Q)\;,
\label{eq:onelooptad}
\end{equation}
where $t_i$, with $i=d,u$, are the renormalized tadpoles, $t_i^0$ are given in
(\ref{eq:tadpoles}), and $\tilde T_i(Q)$ are the renormalized one--loop
contributions at the scale $Q$. Here we neglected the one--loop corrections for
$t_3$ since we are only interested in the value of $\mu$.

\begin{figure}
\centerline{\protect\hbox{\psfig{file=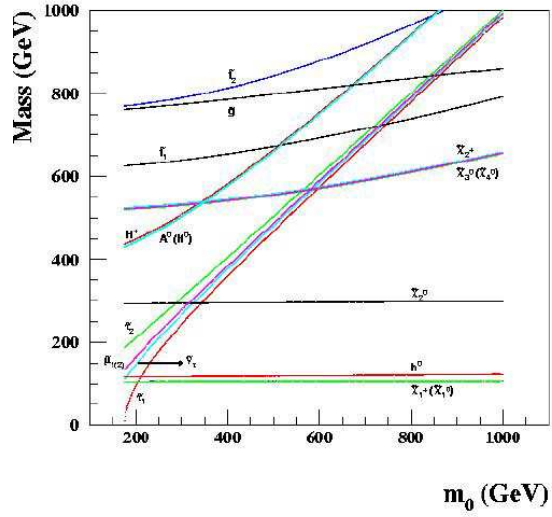,height=11cm}}}
\caption{  Supersymmetric mass spectrum in AMSB--BRpV for $m_{3/2}=32$ TeV,
  $\tan\beta=5$, and $\mu<0$. The values of $\epsilon_3$ and $m_{\nu_{\tau}}$
  were randomly varied according to $10^{-5}<\epsilon_3<1$ GeV and
  $10^{-6}<m_{\nu_{\tau}}<1$ eV. }
\label{masses_32000_5}
\end{figure}

Using the procedure underlined above, the whole mass spectrum can be
calculated as a function of the input parameters $m_0$, $m_{3/2}$,
$\tan\beta$, ${\mathrm{sign}}(\mu)$, $\epsilon_3$, and $m_{\nu_{\tau}}$. In
Fig.~\ref{masses_32000_5}, we show a scatter plot of the mass spectrum as a
function of the scalar mass $m_0$ for $m_{3/2}=32$ TeV, $\tan\beta=5$, and
$\mu<0$, varying $\epsilon_3$ and $m_{\nu_{\tau}}$ according to
$10^{-5}<\epsilon_3<1$ GeV and $10^{-6}<m_{\nu_{\tau}}<1$ eV.  The widths of
the scatter plots show that the spectrum exhibits a very small dependence on
$\epsilon_3$ and $m_{\nu_{\tau}}$. Throughout this paper we use this range for
$\epsilon_3$ and $m_{\nu_{\tau}}$ in all figures.

We can see from this figure that, for $m_0\gsim 200$ GeV, the LSP is the
lightest neutralino $\tilde{\chi}^0_1$ with the lightest chargino
$\tilde{\chi}^+_1$ almost degenerated with it, as in $R_P$--conserving AMSB.
Nevertheless, the LSP is the lightest stau $\tilde{\tau}_1^+$ for $m_0\lsim
200$ GeV.  This last region of parameter space is forbidden in
$R_P$--conserving AMSB, but perfectly possible in AMSB--BRpV since the stau is
unstable, decaying into $R_P$--violating modes with sizeable branching
ratios. Furthermore, the slepton masses have a strong dependence on $m_0$. We
plotted masses of the two staus, which have an appreciable splitting, the
almost degenerated smuons, and the closely degenerated
tau--sneutrinos\footnote{In fact, there are two tau--sneutrinos in
this model, a CP--even and a CP--odd field that are almost degenerated; 
see further Sections for details.}.  
The heavy Higgs bosons have also a strong dependence on $m_0$
and, for the chosen parameters, they are much heavier than the sleptons. On
the other hand, the gauginos show little dependence on $m_0$, as expected.
\vskip 1.5cm

\section{CP--odd Higgs/Sneutrino Sector}
\label{cpodd}
\vskip .9cm

In our model, the CP--odd Higgs sector mixes with the imaginary part of the
tau--sneutrino due to the bilinear $R_P$ violating interactions. Writing the
mass terms in the form
\begin{equation} 
V_{quadratic}=\half[\varphi^0_d,\varphi^0_u,\tilde\nu_{\tau}^{i0}] 
{\bold M^2_{P^0}}\left[\matrix{ 
\varphi^0_d \cr \varphi^0_u \cr \tilde\nu_{\tau}^{i0}
}\right]\;,
\label{NeutOddScalLag} 
\end{equation} 
we have
\begin{equation}
{\scriptsize
{\bold M_{P^0}^2}= 
\left[\matrix{
m_A^{2(0)}s_{\beta}^2+\mu\epsilon_3{{v_3}\over{v_d}}
& m_A^{2(0)}s_{\beta}c_{\beta}
& -\mu\epsilon_3
\cr m_A^{2(0)}s_{\beta}c_{\beta}
& m_A^{2(0)}c_{\beta}^2-\mu\epsilon_3{{v_3}\over{v_d}}
{{c_{\beta}^2}\over{s_{\beta}^2}}+{{v_3^2}\over{v_d^2}}
{{c_{\beta}^2}\over{s_{\beta}^2}}\overline m_{\tilde\nu_{\tau}}^2
& -\mu\epsilon_3{{c_{\beta}}\over{s_{\beta}}}+
{{v_3}\over{v_d}}{{c_{\beta}}\over{s_{\beta}}}\overline 
m_{\tilde\nu_{\tau}}^2
\cr -\mu\epsilon_3
& -\mu\epsilon_3{{c_{\beta}}\over{s_{\beta}}}+
{{v_3}\over{v_d}}{{c_{\beta}}\over{s_{\beta}}}\overline 
m_{\tilde\nu_{\tau}}^2
& \overline m_{\tilde\nu_{\tau}}^2
}\right]\;,
}
\label{OddScalM0}
\end{equation}
with $\overline m_{\tilde\nu_{\tau}}^2=m_{\tilde\nu_{\tau}}^{2(0)}+
\epsilon_3^2+\eighth g_Z^2v_3^2$ and  $g_Z^2\equiv g^2+g'^2$. Here,
\begin{equation}
m_A^{2(0)}={{B\mu}\over{s_{\beta}c_{\beta}}} \qquad \hbox{and} \qquad
m_{\tilde\nu_{\tau}}^{2(0)}=M_{L_3}^2+\eighth g_Z^2(v_d^2-v_u^2)
\end{equation}
are respectively the CP--odd Higgs and sneutrino masses in the $R_P$ conserving
limit ($\epsilon_3=v_3=0$).  In order to write this mass matrix we have
eliminated $m_{H_u}^2$, $m_{H_d}^2$, and $B_3$ using the tadpole equations
(\ref{eq:tadpoles}). The mass matrix has an explicitly
vanishing eigenvalue, which corresponds to the neutral Goldstone boson.

This matrix can be diagonalized with a rotation
\begin{equation} 
\left[\matrix{A^0 \cr G^0 \cr \tilde\nu_{\tau}^{odd}}\right]= 
{\bold R_{P^0}}\left[\matrix{ 
\varphi^0_d \cr \varphi^0_u \cr \tilde\nu_{\tau}^{i0}}\right]\;, 
\label{eigenvCPodd} 
\end{equation} 
where $G^0$ is the massless neutral Goldstone boson. Between the other two
eigenstates, the one with largest $ \tilde\nu_{\tau}^{i0}$ component is called
CP--odd tau--sneutrino $\tilde\nu_{\tau}^{odd}$ and the remaining state is
called CP--odd Higgs $A^0$.

As an intermediate step, it is convenient to make explicit the masslessness 
of the Goldstone boson with the rotation
\begin{equation}
{\bold{\widehat R_{P^0}}}=
\left[\matrix{
s_{\beta} & c_{\beta} & 0 \cr
-c_{\beta}r & s_{\beta}r & -{{v_3}\over{v_d}}c_{\beta}r \cr
-{{v_3}\over{v_d}}c_{\beta}^2r & {{v_3}\over{v_d}}s_{\beta}c_{\beta}r & r
}\right]\;,
\end{equation}
where 
\begin{equation}
r={1\over{\sqrt{1+{{v_3^2}\over{v_d^2}}c_{\beta}^2}}}\;,
\label{ar}
\end{equation}
obtaining a rotated mass matrix $\widehat R_{P^0} M_{P^0}^2 \widehat
R_{P^0}^{T}$ which has a column and a row of zeros, corresponding to $G^0$.
This procedure simplifies the analysis since the remaining $2\times2$ mass
matrix for $(A^0,\tilde\nu_{\tau}^{odd})$ is
\begin{equation}
{\bold{\widehat M}}^2_{P^0}=
\left[\matrix{
m_A^{2(0)}+{{v_3^2}\over{v_d^2}}{{c_{\beta}^4}\over{s_{\beta}^2}}
\overline m_{\tilde\nu_{\tau}}^2+\mu\epsilon_3{{v_3}\over{v_d}}
{{s_{\beta}^2-c_{\beta}^2}\over{s_{\beta}^2}}
& \left({{v_3}\over{v_d}}{{c_{\beta}^2}\over{s_{\beta}}}
\overline m_{\tilde\nu_{\tau}}^2-\mu\epsilon_3{1\over{s_{\beta}}}\right)r
\cr \left({{v_3}\over{v_d}}{{c_{\beta}^2}\over{s_{\beta}}}\overline 
m_{\tilde\nu_{\tau}}^2-\mu\epsilon_3{1\over{s_{\beta}}}\right)r
& \overline m_{\tilde\nu_{\tau}}^2
{1\over r^2}
}\right]\;.
\label{OddScalM0rot}
\end{equation}
%


%
\begin{figure}
\centerline{\protect\hbox{\psfig{file=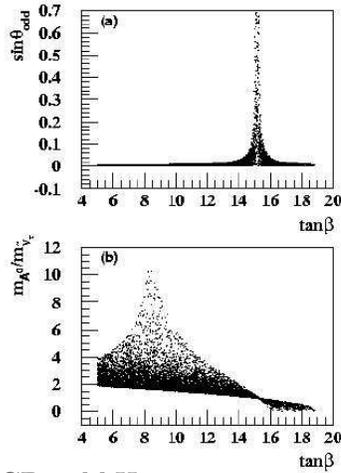,height=11cm}}}
\vskip -1.5cm
\caption{  (a) CP--odd Higgs--sneutrino mixing and (b) ratio between the 
  CP--odd Higgs mass and the sneutrino mass as a function of $\tan \beta$ for
  $m_{3/2}=32$ TeV, $\mu<0$ and $100 < m_0 < 300$ GeV.  }
\label{snua_tgb}
\end{figure}

We quantify  the mixing between the tau--sneutrino and the neutral Higgs bosons
through
\begin{equation}
\sin^2{\theta_{odd}} = |\langle\tilde\nu_{\tau}^{odd}|\varphi^0_u\rangle|^2+
|\langle\tilde\nu_{\tau}^{odd}|\varphi^0_d\rangle|^2 \; .
\label{snutau_higgs}
\end{equation}
If we consider the $R_P$ violating interactions as a perturbation, we can show
that 
\begin{equation}
\sin^2{\theta_{odd}} \simeq{{\left(
{{v_3}\over{v_d}}c_{\beta}^2m_{\tilde\nu_{\tau}}^{2(0)}-\mu\epsilon_3
\right)^2}\over{
s_{\beta}^2\left(m_A^{2(0)}-m_{\tilde\nu_{\tau}}^{2(0)}\right)^2
}} + \frac{v_3^2}{v_d^2}c^2_\beta \; ,
\end{equation}
indicating that this mixing can be large when the CP--odd Higgs boson $A^0$ and
the sneutrino $\tilde\nu_{\tau}$ are approximately degenerate.

Figure \ref{snua_tgb}a displays the full sneutrino--Higgs mixing
(\ref{snutau_higgs}), with no approximations, 
as a function of $\tan\beta$ for $m_{3/2} = 32$ TeV, $\mu
< 0$ and $100 < m_0 < 300$ GeV.  In a large fraction of the parameter
space this mixing is small, since it is proportional to the BRpV parameters
squared divided by MSSM mass parameters squared. 
However, it is possible to find
a region where the mixing is sizable, {\em e.g.}, for our choice of parameters
this happens at $\tan\beta\approx 15$. As expected, the region of large
mixing is associated to near degenerate states, as we can see from
Fig.~\ref{snua_tgb}b where we present the ratio between the CP--odd Higgs mass
$m_{A}$ and the CP--odd tau--sneutrino mass $m_{\tilde\nu_{\tau}^{odd}}$ as a
function of $\tan\beta$.


\section{CP--even Higgs/Sneutrino Sector}
\label{cpeven}

The mass terms of the CP--even neutral scalar sector are
\begin{equation} 
V_{quadratic}=\half[\chi^0_d,\chi^0_u,\tilde\nu_{\tau}^{r0}] 
{\bold M^2_{S^0}}\left[\matrix{ 
\chi^0_d \cr \chi^0_u \cr \tilde\nu_{\tau}^{r0}
}\right]\;,
\label{NeutEvenScalLag} 
\end{equation} 
where the mass matrix can be separated into two pieces
\begin{equation}
{\bold M^2_{S^0}}={\bold M_{S^0}^{2(0)}}+{\bold M_{S^0}^{2(1)}} \;.
\end{equation}
The first term due to $R_P$ conserving interactions is
\begin{equation}
{\bold M_{S^0}^{2(0)}}= 
\left[\matrix{
m_A^{2(0)}s_{\beta}^2+\fourth g_Z^2v_d^2
& -m_A^{2(0)}s_{\beta}c_{\beta}-\quarter g^2_Zv_dv_u  
& 0
\cr -m_A^{2(0)}s_{\beta}c_{\beta}-\quarter g^2_Zv_dv_u  
& m_A^{2(0)}c_{\beta}^2+\fourth g_Z^2v_u^2
& 0
\cr 0
& 0
& m_{\tilde\nu_{\tau}}^{2(0)}
}\right]\;,
\label{EvenScalM0}
\end{equation}
while the one associated to the $R_P$ violating terms is
\begin{equation}
{\scriptsize
\!\!{\bold M_{S^0}^{2(1)}}= 
\left[\matrix{
\mu\epsilon_3{{v_3}\over{v_d}}
& 0
& -\mu\epsilon_3+\quarter g^2_Zv_dv_3  
\cr 0
& {{v_3^2}\over{v_d^2}}{{c_{\beta}^2}\over{s_{\beta}^2}}
m_{\tilde\nu_{\tau}}^{2(0)}-
\mu\epsilon_3{{v_3}\over{v_d}}{{c_{\beta}^2}\over{s_{\beta}^2}}
& \mu\epsilon_3{{c_{\beta}}\over{s_{\beta}}}-
{{v_3}\over{v_d}}{{c_{\beta}}\over{s_{\beta}}}m_{\tilde\nu_{\tau}}^{2(0)}
-\quarter g^2_Zv_uv_3
\cr -\mu\epsilon_3+\quarter g^2_Zv_dv_3  
& \mu\epsilon_3{{c_{\beta}}\over{s_{\beta}}}-
{{v_3}\over{v_d}}{{c_{\beta}}\over{s_{\beta}}}m_{\tilde\nu_{\tau}}^{2(0)}
-\quarter g^2_Zv_uv_3
& \epsilon_3^2+{\textstyle{3\over8}}g^2_Zv_3^2
}\right]\;.
}
\label{EvenScalM1}
\end{equation}
Radiative corrections can change significantly the lightest Higgs mass
and, consequently, we have also introduced the leading correction to its
mass

\begin{equation}
\Delta m_{\chi^0_u} \equiv \frac{3m^4_t}{2\pi^2v_u^2v^{\prime}}\ln
\left(\frac{m_{\tilde{t}_1}m_{\tilde{t}_2}}{m^2_t}\right)\;,
\label{HiggsCorr} 
\end{equation}
with 
\begin{equation}
v^{\prime} = 1 - \frac{v_3^2}{v_d^2+v_u^2+v_3^2}\;,
\end{equation}
by adding it to the element $[{\bold M_{S^0}^{2}}]_{22}$.

Analogously to the CP--odd sector, we define the mixing between the 
CP--even tau--sneutrino and the neutral Higgs bosons as 
\begin{equation}
\sin^2{\theta_{even}}=|\langle\tilde\nu_{\tau}^{even}|\chi^0_d\rangle|^2+
|\langle\tilde\nu_{\tau}^{even}|\chi^0_u\rangle|^2=
|\langle H^0|\tilde\nu_{\tau}^{r0}\rangle|^2+
|\langle h^0|\tilde\nu_{\tau}^{r0}\rangle|^2\;.
\label{snutau_higgs2}
\end{equation}
In general, this mixing is small since it is proportional to the $R_P$
breaking parameters squared, however, it can be large provided the
sneutrino is degenerate either with $h^0$ or $H^0$.

\begin{figure}
\centerline{\protect\hbox{\psfig{file=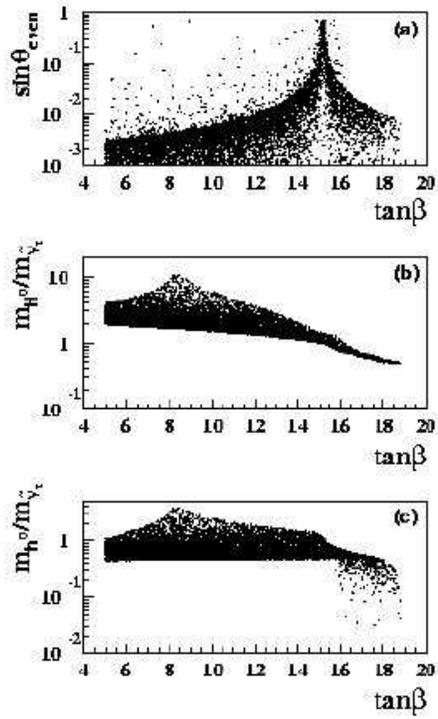,height=17cm}}}
\vskip -1cm
\caption{
  (a) CP--even Higgs--sneutrino mixing; (b) ratio between heavy CP--even Higgs
  and tau--sneutrino masses and (c) ratio between light CP--even Higgs and
  tau--sneutrino masses as a function of $\tan\beta$ for $m_{3/2}=32$ TeV,
  $\mu<0$ and $100 < m_0 < 300$ GeV.}
\label{snuh_tgb}
\end{figure}

In Figure \ref{snuh_tgb}a, we present the mixing (\ref{snutau_higgs2}) as a
function of $\tan\beta$, for the input parameters as in Fig.~\ref{snua_tgb}.
Similarly to the CP--odd scalar sector, this mixing can be very large,
occurring either when $m_H\approx m_{\tilde\nu_{\tau}^{even}}$ or $m_h\approx
m_{\tilde\nu_{\tau}^{even}}$. In fact, we can see from Fig.~\ref{snuh_tgb}b
that the peak in Fig.~\ref{snuh_tgb}a for $\tan\beta\sim 15$ is mainly due to
the mass degeneracy between the heavy CP--even Higgs $H^0$ and the CP--even
tau--sneutrino $\tilde\nu_{\tau}^{even}$. On the other hand, the other
scattered dots with high mixing angle values throughout Fig.~\ref{snuh_tgb}a
come from points in the parameter space where the light CP--even Higgs $h^0$
and the CP--even tau--sneutrino $\tilde\nu_{\tau}^{even}$ are degenerated. We
see from Fig.~\ref{snuh_tgb}c that this may occur for $5 < \tan\beta < 15$.


It is important to notice that the enhancement of the mixing between the
tau--sneutrino and the CP--even Higgs bosons for almost degenerate states
implies that large $R_P$ violating effects are possible even for small $R_P$
violating parameters ($\epsilon_3\lsim 1$ GeV), and for neutrino masses
consistent with the solutions to the atmospheric neutrino anomaly
($m_{\nu_{\tau}}\lsim 1$ eV).

\vskip 1.5cm
\section{Charged Higgs/Charged Slepton Sector}
\label{charged}
\vskip 1cm

The mass terms in the charged scalar sector are
\begin{equation} 
V_{quadratic}=[H_u^-,H_d^-,\tilde\tau_L^-,\tilde\tau_R^-] 
{\bold M^2_{S^{\pm}}}\left[\matrix{ 
H_u^+ \cr H_d^+ \cr \tilde\tau_L^+ \cr \tilde\tau_R^+
}\right]\;,
\label{ChaScalLag} 
\end{equation} 
where it is convenient to split the mass matrix into a $R_P$ conserving part
and a $R_P$ violating one.
\begin{equation}
{\bold M^2_{S^{\pm}}}={\bold M^{2(0)}_{S^{\pm}}}+{\bold M^{2(1)}_{S^{\pm}}}\;.
\end{equation}
The $R_P$ conserving mass matrix has the usual MSSM form
%

\noindent${\small
{\bold M^{2(0)}_{S^{\pm}}}=
}
$
\begin{equation}
{\scriptsize
=\left[\matrix{
m_A^{2(0)}s_{\beta}^2+\fourth g^2v_u^2 & 
m_A^{2(0)}s_{\beta}c_{\beta}+\fourth g^2v_uv_d & 
0 & 0
\cr
m_A^{2(0)}s_{\beta}c_{\beta}+\fourth g^2v_uv_d & 
m_A^{2(0)}c_{\beta}^2+\fourth g^2v_d^2 & 
0 & 0
\cr
0 & 0 & 
\widehat M_{L_3}^2 & {1\over{\sqrt{2}}}h_{\tau}(A_{\tau}v_d-\mu v_u)
\cr
0 & 0 & 
{1\over{\sqrt{2}}}h_{\tau}(A_{\tau}v_d-\mu v_u) & \widehat M_{R_3}^2
}\right]\;,
}
\label{ch_higgs_matrix}
\end{equation}
where $h_\tau$ is the $\tau$ Yukawa coupling and
\begin{eqnarray}
\widehat M_{L_3}^2&=&
M_{L_3}^2-\eighth(g^2-g'^2)(v_d^2-v_u^2)+\half h_{\tau}^2v_d^2\;,
\nonumber\\
\widehat M_{R_3}^2&=&
M_{R_3}^2-\fourth g'^2(v_d^2-v_u^2)+\half h_{\tau}^2v_d^2\;.
\end{eqnarray}

The contribution due to $R_P$ violating terms is

\noindent${\small
{\bold M^{2(1)}_{S^{\pm}}}=}$
\begin{equation}
{\scriptsize
=\left[\matrix{
\mu\epsilon_3{{v_3}\over{v_d}}-\fourth g^2v_3^2+\half h_{\tau}^2v_3^2 & 
0 
& X_{uL} & X_{uR}
\cr
0 & 
{{v_3^2}\over{v_d^2}}{{c_{\beta}^2}\over{s_{\beta}^2}}
\overline{m}_{\tilde\nu}^2-\mu\epsilon_3{{v_3}\over{v_d}}
{{c_{\beta}^2}\over{s_{\beta}^2}}+\fourth g^2v_3^2 
& X_{dL} & X_{dR}
\cr
X_{uL} & X_{dL}
& \epsilon_3^2+\eighth g_Z^2v_3^2
& 0
\cr
X_{uR} & X_{dR}
& 0 & \half h_{\tau}^2v_3^2-\fourth g'^2v_3^2
}\right]\;,
}
\end{equation}
with
\begin{eqnarray}
X_{uL} &=& \fourth g^2v_dv_3-\mu\epsilon_3-\half h_{\tau}^2v_dv_3\;,
\label{xul}\\
X_{uR} &=& -{1\over{\sqrt{2}}}h_{\tau}(A_{\tau}v_3+\epsilon_3v_u)\;,
\label{xur}\\
X_{dL} &=& {{v_3}\over{v_d}}{{c_{\beta}}\over{s_{\beta}}}
\overline{m}_{\tilde\nu}^2-\mu\epsilon_3
{{c_{\beta}}\over{s_{\beta}}}+\fourth g^2v_uv_3\;,
\label{xdl}\\
X_{dR} &=& -{1\over{\sqrt{2}}}h_{\tau}(\mu v_3+\epsilon_3v_d)\;.
\label{xdr}
\end{eqnarray}

The complete matrix ${\bold M^2_{S^{\pm}}}$ has an explicit zero eigenvalue
corresponding to the charged Goldstone boson $G^{\pm}$, and is diagonalized by
a rotation matrix ${\bold R_{S^{\pm}}}$ such that
\begin{equation}
\left[\matrix{ 
H^+ \cr G^+ \cr \tilde\tau_1^+ \cr \tilde\tau_2^+
}\right]
={\bold R_{S^{\pm}}}
\left[\matrix{ 
H_u^+ \cr H_d^+ \cr \tilde\tau_L^+ \cr \tilde\tau_R^+
}\right]\;.
\end{equation}

In analogy with the discussion on the CP--even scalar sector, we define the
mixing of the lightest (heaviest) stau $\tilde\tau_1^\pm$ ($\tilde\tau_2^\pm$)
with the charged Higgs bosons as 
\begin{eqnarray}
\sin^2{\theta^+_1}&=&|\langle\tilde\tau_1^+|H^+_u\rangle|^2+
|\langle\tilde\tau_1^+|H^+_d\rangle|^2 \;,\\
\sin^2{\theta^+_2}&=&|\langle\tilde\tau_2^+|H^+_u\rangle|^2+
|\langle\tilde\tau_2^+|H^+_d\rangle|^2 \;.
\label{stauHigsmix}
\end{eqnarray}
\begin{figure}
\centerline{\protect\hbox{\psfig{file=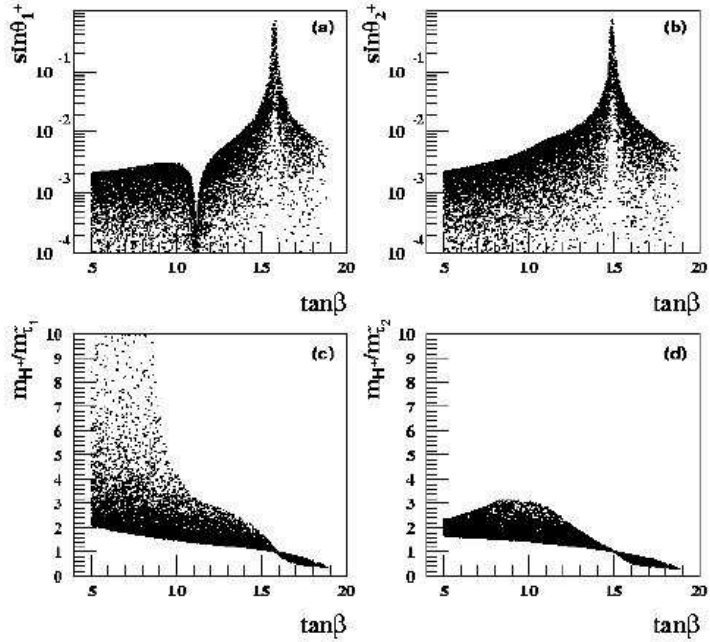,height=15.cm}}}
\caption{
  (a) Charged Higgs--light stau mixing; (b) Charged Higgs--heavy stau 
  mixing; (c) charged Higgs--light stau mass ratio and (d) charged 
  Higgs--heavy stau mass
  ratio as a function of $\tan\beta$ for $m_{3/2}=32$ TeV, $\mu<0$ and $100 <
  m_0 < 300$ GeV.}
\label{stau_sch}
\end{figure}

Figure \ref{stau_sch}a~(b) contains the mixing between the
lightest~(heaviest) stau and the
charged Higgs fields $\sin\theta^+_{1(2)}$ as a function of $\tan\beta$ for
$m_{3/2}=32$ TeV, $\mu<0$, and $100<m_0<300$ GeV. In this sector, the mixing
can also be very large provided there is a near degeneracy between the staus
$\tilde{\tau}^{\pm}_1$, $\tilde{\tau}^{\pm}_2$ and $H^{\pm}$. 
We can see clearly this effect in
Fig.~\ref{stau_sch}c~(d), where we show the ratio between the charged 
Higgs mass $m_{H^+}$ and the lightest~(heaviest) stau mass 
$m_{\tilde{\tau}_{1(2)}}$. In
Figs.~\ref{stau_sch}a and b we also notice that large 
light stau--charged
Higgs mixing occurs at slight different value of $\tan\beta$ compared
with heavy stau--charged Higgs mixing. Large light stau--charged Higgs
mixing is found in Fig.~\ref{stau_sch}a as a peak at
$\tan\beta \approx 16$, as opposed to large heavy stau--charged Higgs
mixing, which presents a peak at $\tan\beta \approx 15$. 
In Fig.~\ref{stau_sch}a we notice that the mixing angle vanishes 
at $\tan\beta\sim 11$. This zero occurs at the point of parameter
space where the two staus are nearly degenerated, as will be explained in 
Sec.~\ref{discussion}.

\begin{figure}[htb]
\centerline{\protect\hbox{\psfig{file=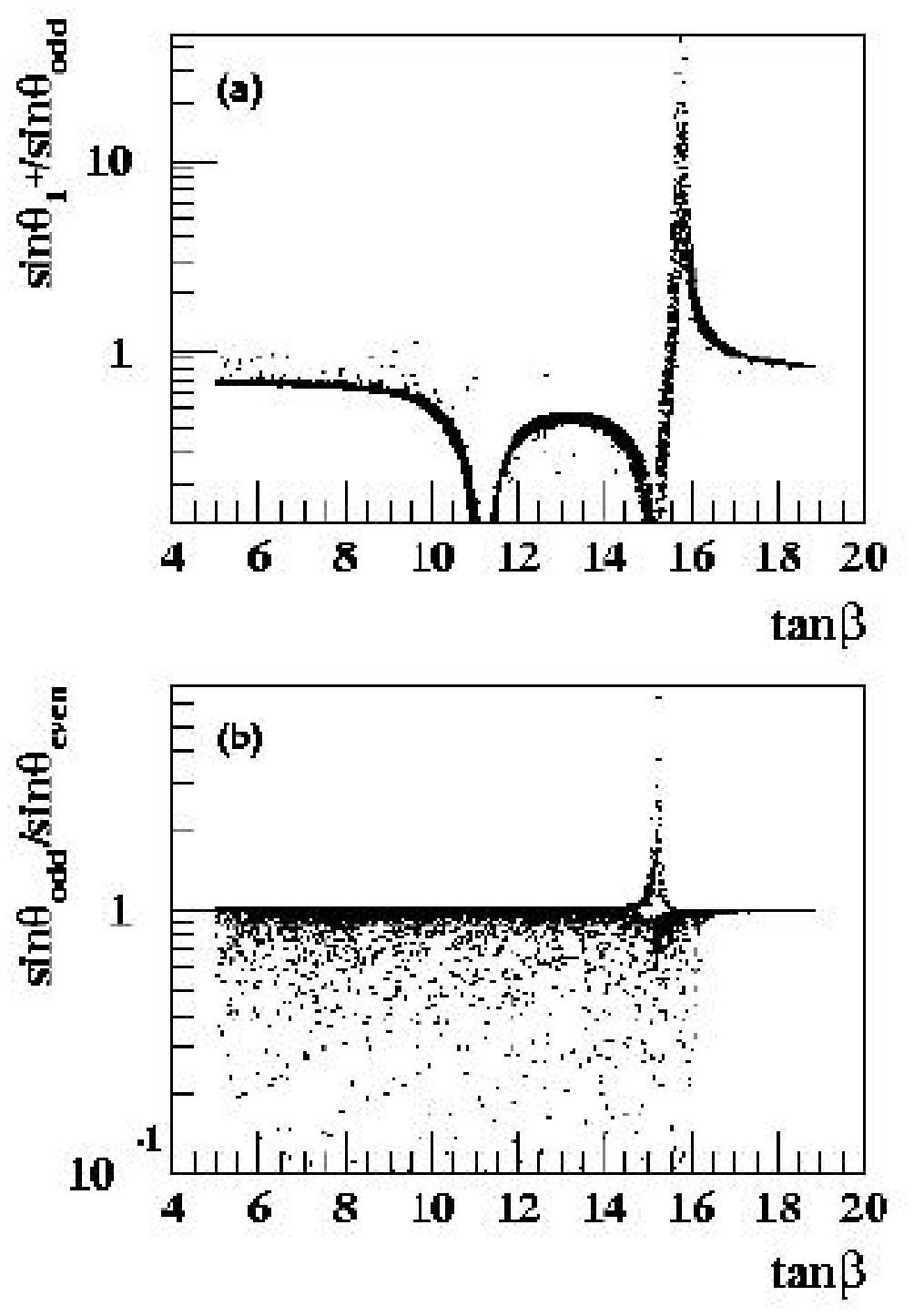,height=10.6cm}}}
\caption{
  (a) Ratio between the charged Higgs--stau and CP--odd Higgs--tau--sneutrino
  mixing angles and (b) ratio between the CP--odd Higgs--tau--sneutrino and
 CP--even Higgs--tau--sneutrino mixing angles as a function of $\tan\beta$ for
  $m_{3/2}=32$ TeV, $\mu<0$ and $100<m_0<300$ GeV.  }
\label{ratio}
\end{figure}

Similarly, in the last figure, the exact value of $\tan\beta$ 
at which the peak of the lightest stau--charged scalar mixing occurs 
is somewhat larger than the analogous
mixing for the CP--odd sector $\sin\theta_{odd}$. 
This can be appreciated in Fig.~\ref{ratio}a
where we show the ratio between $\sin \theta_1^+$ and $\sin \theta_{odd}$ as a
function of $\tan\beta$ for $m_{3/2}=32$ TeV, $\mu<0$ and $100<m_0<300$ GeV.
The peak of the charged sector mixing is located at the peak of the ratio. On
the other hand, the peak for the neutral CP--odd sector is located at the
nearby zero of the ratio. The other zero of the ratio near $\tan\beta\approx
11$ corresponds to a zero of the charged scalar sector mixing, as shown in
Fig.~\ref{stau_sch}.  
For the sake of comparison, we display in
Fig.~\ref{ratio}b the ratio between the CP--odd and CP--even mixings
($\sin\theta_{odd}/\sin\theta_{even}$) as a function of $\tan\beta$. We can
see that most of the time the ratio is equal to 1 showing that the two neutral
scalar sectors have similar behavior with $\tan\beta$ in contrast with the
charged scalar sector.  The points where this ratio is lower than 1 correspond
to the case where the CP--even scalar sector mixings are dominated by the light
Higgs and tau--sneutrino degeneracy which occurs for any value of $\tan\beta$
lower than 16, as shown in Fig.~\ref{snuh_tgb}c.

\section{The Neutrino Mass}
\label{neutrino}

BRpV provides a solution to the atmospheric and solar neutrino problems 
due to their mixing with neutralinos, which generates neutrino masses 
and mixing angles. 
It was shown in \cite{numassBRpV} that the atmospheric mass scale is 
adequately described by the tree level neutrino mass
\begin{equation}
m_{\nu_3}^{tree}={{M_1g^2+M_2g'^2}\over{4\Delta_0}}|\vec\Lambda|^2\;,
\label{nutree}
\end{equation}
where $\Delta_0$ is the determinant of the neutralino sub--matrix and
$\vec\Lambda=(\Lambda_1,\Lambda_2,\Lambda_3)$, with
\begin{equation}
\Lambda_i=\mu v_i+\epsilon_i v_d\;,
\label{lambda}
\end{equation}
where the index $i$ refers to the lepton family. The spectrum generated is 
hierarchical, and obtained typically with 
$\Lambda_1\ll\Lambda_2\approx\Lambda_3$.

\begin{figure}[thb]
\centerline{\protect\hbox{\psfig{file=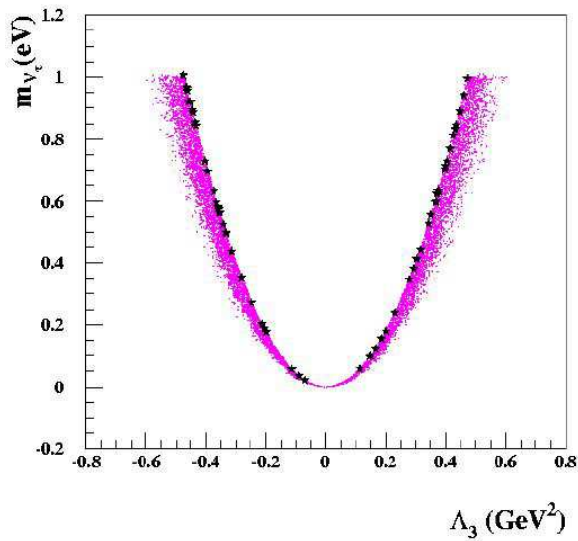,height=11.5cm}}}
\caption{
Tau neutrino mass as a function of $\Lambda_3$ for $5 < \tan \beta < 20$, 
$100<m_0< 1000$ GeV, $m_{3/2}=32$ TeV and $\mu < 0$. 
}
\label{nu_lambda}
\end{figure}

As it was mentioned in the introduction, for many purposes it is enough to
work with $R_P$ violation only in the third generation. In this case, the
atmospheric mass scale is well described by Eq.~(\ref{nutree}) with the
replacement $|\vec\Lambda|^2\rightarrow\Lambda_3^2$. In Fig.~\ref{nu_lambda},
we plot the neutrino mass as a function of $\Lambda$ in AMSB--BRpV with the
input parameters $m_{3/2}=32$ TeV, $\mu<0$, $5<\tan\beta<20$, $100<m_0<1000$
GeV and $10^{-5}<\epsilon_3<1$ GeV. The
quadratic dependence of the neutrino mass on $\Lambda$ is apparent in this
figure and neutrino masses smaller than 1 eV occur for $|\Lambda|\lsim0.6\,
{\mathrm{GeV}}^2$. Moreover, the stars correspond to the allowed neutrino
masses when the tau--sneutrino is the LSP. In general the points with a small
(large) $m_0$ are located in the inner (outer) regions of this scattered plot.

From Fig.~\ref{nu_lambda}, we can see that the attainable neutrino masses are
consistent with the global three--neutrino oscillation data analysis in the
first reference of \cite{NeuAnom} that favors the $\nu_{\tau}\rightarrow
\nu_{\mu}$ oscillation hypothesis.  Although only mass squared differences are
constrained by the neutrino data, our model naturally gives a hierarchical
neutrino mass spectrum, therefore, we extract a na\"{\i}ve constraint on the
actual mass coming from the analysis of the full atmospheric neutrino data,
$0.04\lsim m_{\nu_{\tau}}\lsim 0.09$ eV \cite{NeuAnom}.  In addition, we
notice that it is not possible to find an upper bound on the neutrino mass if
angular dependence on the neutrino data is not included and only
the total event rates are considered. 

\begin{figure}[thb]
\centerline{\protect\hbox{\psfig{file=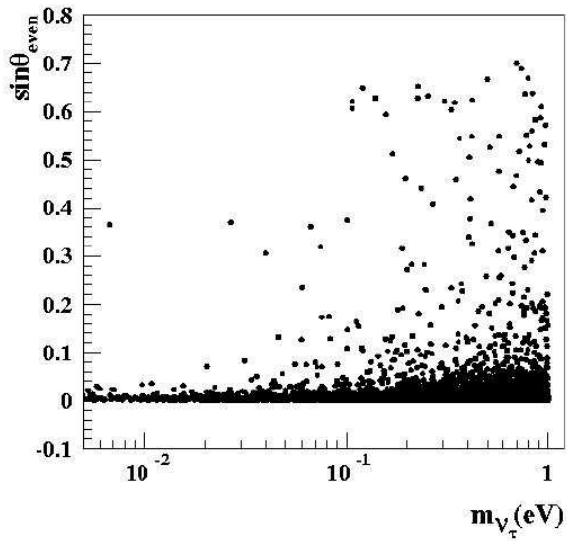,height=11.5cm}}}
\caption{
  Mixing between CP--even Higgses and sneutrino as a function 
  of the tau neutrino mass.}
\label{nuh_nutau}
\end{figure}

In Fig.~\ref{nuh_nutau} we show the correlation between the neutrino mass and
mixing of the tau--sneutrino and the CP--even Higgses 
($\sin\theta_{even}$) for
the parameters assumed in Fig.~\ref{nu_lambda}. As expected, the largest
mixings are associated to larger neutrino masses. Notwithstanding, it is
possible to obtain large mixings for rather small neutrino masses because the
mixing is proportional to the $R_P$ violating parameters $\epsilon_3$ and
$v_3$, and not directly on $\Lambda_3 \propto m_{\nu_\tau}$.
In any case, Fig.~\ref{nuh_nutau} suggests that large
scalar mixings are still possible even imposing these bounds on the neutrino
mass. This is extremely important for the phenomenology of the model because
it indicates that non negligible $R_P$ violating branching ratios are possible
for scalars even in the case they are not the LSP.


\section{Discussions}
\label{discussion}

\begin{figure}[htb]
\centerline{\protect\hbox{\psfig{file=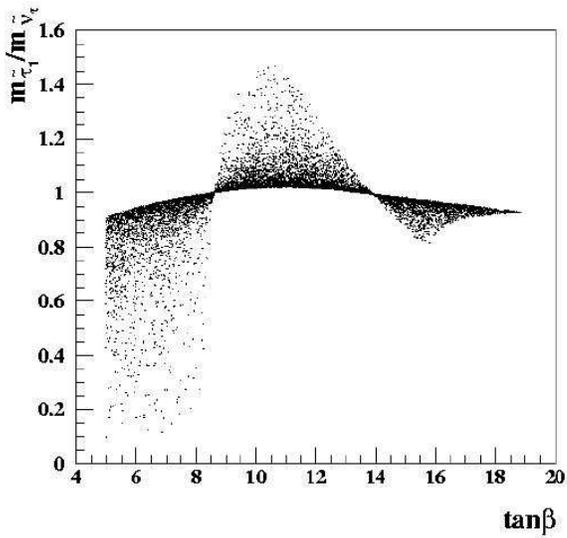,height=11.5cm}}}
\caption{ 
   Ratio between the light stau and the sneutrino masses as a
   function of $\tan \beta$ for $m_{3/2}=32$ TeV, $100<m_0<300$ GeV
   and $\mu<0$.}
\label{stau_snu}
\end{figure}

The presence of $R_P$ violating interactions in our model render the
LSP unstable, avoiding strong constraints on the possible LSP
candidates.  In the parameter regions where the neutralino is not the
LSP, whether the light stau or the tau--sneutrino is the LSP depends
crucially on the value of $\tan\beta$. This fact can be seen in
Fig.~\ref{stau_snu} where we plot the ratio between the light stau and
the tau--sneutrino masses as a function of $\tan\beta$ for $m_{3/2}=32$
TeV, $100<m_0<300$ GeV, and $\mu<0$.  From this figure we see that the
tau--sneutrino is the LSP for $8.5 \lsim \tan \beta \lsim 14$,
otherwise the stau is the LSP\footnote{Of course, $m_0$ has
to be small enough so that the slepton is lighter than the
neutralino.}.

When the stau is the LSP, it decays via $R_P$ violating interactions,
\ie, its decays take place through mixing  with the charged Higgs, and
consequently, they will mimic the charged Higgs boson ones. Therefore,
it is very important to be able to distinguish between $\tilde\tau^{\pm}_1$
and $H^{\pm}$. This can be achieved either through precise studies of
branching ratios, or via the mass spectrum, or both \cite{ALS}.

%
%

\begin{figure}
\centerline{\protect\hbox{\psfig{file=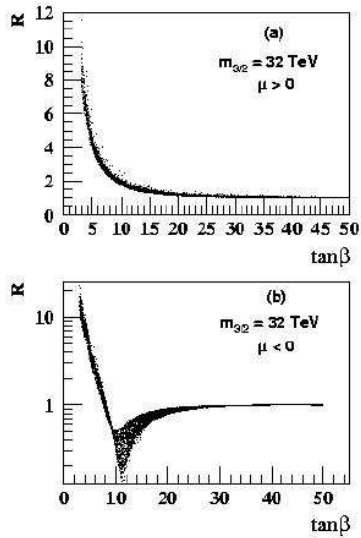,height=12.5cm}}}
\vskip -1.5cm
\caption{
   Ratio ($R$) between the stau splitting in AMSB with and without
   $R_P$ violation as a function of $\tan \beta$, for: $m_{3/2}=32$
   TeV, $100<m_0<1000$ GeV and (a) $\mu>0$ or (b) $\mu<0$.}
\label{rpxmssm_new}
\end{figure}

Measurements on the mass spectrum are also important in order to
distinguish AMSB with and without conservation of $R_P$. In
Fig.~\ref{rpxmssm_new} we present the ratio between the stau
mass splitting in AMSB--BRpV and in the AMSB, $R = (m_{\tilde\tau_2} - 
m_{\tilde\tau_1})_{\rm AMSB-BRpV}/(m_{\tilde\tau_2} - 
m_{\tilde\tau_1})_{\rm AMSB}$, with $\epsilon_3=v_3=0$ and keeping
the rest of the parameters unchanged, as a function of $\tan\beta$.
In these figures, we took $100<m_0<1000$ GeV, $m_{3/2}=32$ TeV, and (a)
$\mu>0$, and (b) $\mu<0$. For $\mu>0$ (Fig.~\ref{rpxmssm_new}a), the
stau mass splitting is always larger in the AMSB--BRpV than in the AMSB
by a factor that increases when $\tan\beta$ decreases, and can be as
large as $R\sim10$ for $\tan\beta\sim3$!  We remind the reader that,
in the absence of $R_P$ violation, the left--right stau mixing
decreases with decreasing $\tan\beta$, thus augmenting the importance
of $R$--parity violating mixings. On the other hand, for $\mu<0$
(Fig.~\ref{rpxmssm_new}b), this ratio can be as large as before at
small $\tan\beta$, but in addition, the splitting can go to zero in
AMSB--BRpV near $\tan\beta\approx 11$, which also constitutes a sharp
difference with the AMSB. For both signs of $\mu$ the ratio goes to
unity at large $\tan\beta$ because the left--right mixing in the AMSB
is proportional to $\tan\beta$ and dominates over any $R_P$ violating
contribution.

\begin{figure}
\centerline{\protect\hbox{\psfig{file=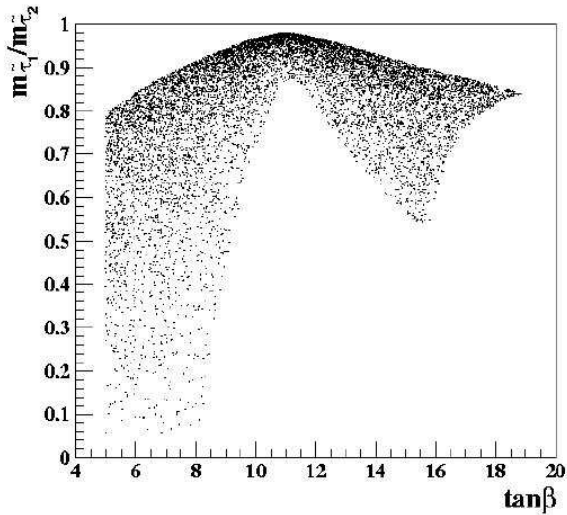,height=11.5cm}}}
\caption{
  Ratio between the light
  and heavy stau masses as a function of $\tan
  \beta$ for $m_{3/2}=32$ TeV,  $100<m_0<300$ GeV and $\mu<0$. 
}
\label{stau}
\end{figure}

The behavior of $R$ at $\tan\beta\sim 11$ in
Fig.~\ref{rpxmssm_new}b indicates that the two staus can be nearly
degenerated in AMSB--BRpV. In Fig.~\ref{stau} we plot the
ratio between the light and heavy stau masses as a function of $\tan\beta$,
for $m_{3/2}=32$ TeV,  $100<m_0<300$ GeV and $\mu<0$, 
observing clearly that the near degeneracy occurs at
$\tan\beta\sim 11$. In first approximation,
consider that the near degeneracy occurs
when $A_\tau v_d-\mu v_u\approx 0$ as inferred from 
Eq.~(\ref{ch_higgs_matrix}). In addition,
the mixing $X_{dR}$ in Eq.~(\ref{xdr}) is also very small because 
it is proportional
to $\Lambda_\tau$ in Eq.~(\ref{lambda}), 
which defines the atmospheric neutrino mass,
as indicated in Eq.~(\ref{nutree}). The smallness of these two
quantities implies that
the mixing $X_{uR}$ in Eq.~(\ref{xur}) is also small in this 
particular region
of parameter space, indicating that the right stau is decoupled from the
Higgs fields and thus originating the zero in the mixing angle, 
noted already in Figs.~\ref{stau_sch} and \ref{ratio}.

\begin{figure}
\centerline{\protect\hbox{\psfig{file=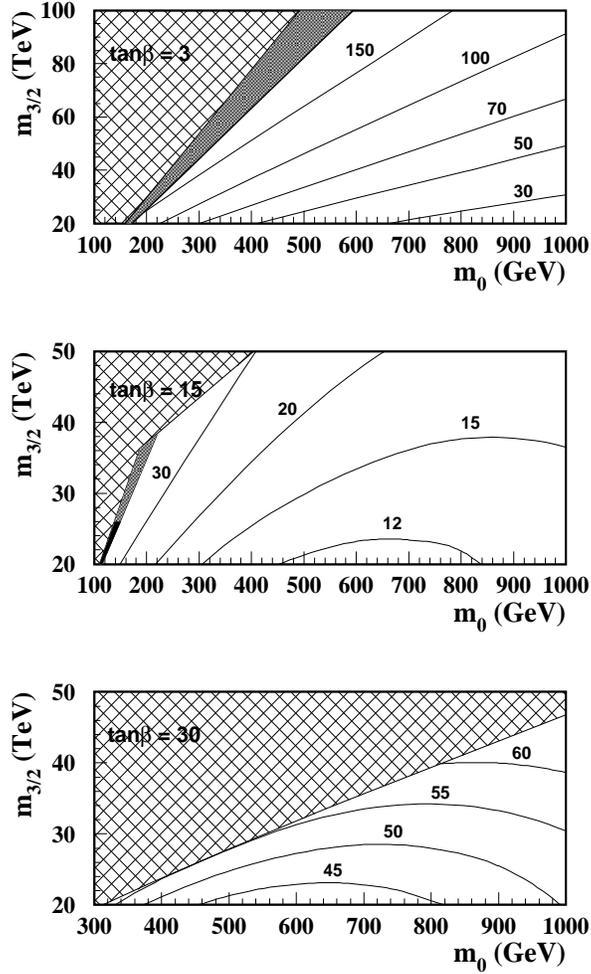,height=17cm}}}
\vskip -1.5cm
\caption{
  Contours of constant splitting between the light stau and heavy stau
  masses in the plane $m_{3/2}\times m_0$ in GeV for $\mu < 0$,
  $\tan\beta = 3$ (a), 15 (b) and 30 (c). The hatched area is
  theoretically forbidden; the grey area in (a) and (b) is where the
  lightest stau is the LSP, while the small black area in (b) is where
  the tau--sneutrino is the LSP.  }
\label{m32_m0}
\end{figure}

In order to quantify the stau mass splitting in our model, we present
in Fig.~\ref{m32_m0} contours of constant splitting between the stau
masses, $m_{\tilde{\tau}_2}-m_{\tilde{\tau}_1}$, in the plane
$m_{3/2}\times m_0$ in GeV for $\mu < 0$ and several $\tan\beta$. We
can see in Fig.~\ref{m32_m0}a that for small $\tan\beta = 3$ the stau
mass splitting in our model starts at $m_{\tilde{\tau}_2} -
m_{\tilde{\tau}_1} \sim 30$ GeV, in sharp contrast with the $R_P$ conserving
case where the biggest splittings barely goes over this value
\cite{fengmoroi}. This is in agreement with the results presented in
Fig.~\ref{rpxmssm_new}b. Furthermore, we can also see that there is a
considerable region in the $m_{3/2}\times m_0$ plane, indicated by the
grey area, where the lightest stau is the LSP.  For intermediary
values of $\tan \beta \sim 15$, Fig.~\ref{m32_m0}b shows that the stau
mass splitting goes to a minimum. This is a different behavior from
the MSSM which presents a mass splitting up to 10 times bigger as we
have seen in Fig.~\ref{rpxmssm_new}b. For this value of $\tan \beta$
we still have a small region where the lightest stau is the LSP (grey
area) and, as a novelty, a tiny region for small values of $m_{3/2}$
and $m_0$ where the tau--sneutrino is the LSP (black area). For large
values of $\tan \beta = 30$, the stau splitting mass shown in
Fig.~\ref{m32_m0}c is similar to the MSSM one
\cite{fengmoroi}.

We have made below a series of three figures fixing the value 
$\tan\beta=15$ to study the dependence on $m_0$ of the mass spectrum and
mixings in the scalar sector. This choice of $\tan\beta$ is
such that we find a degeneracy among the masses, and consequently we
obtain large mixings in the scalar sector. We also chose $m_{3/2}=32$
TeV and $\mu<0$, while the $R_P$ violating parameters were varied
according to $10^{-5}<\epsilon_3<1$ GeV and $10^{-6}<m_{\nu_{\tau}}<1$
eV.

%
%

\begin{figure}[thb]
\centerline{\protect\hbox{\psfig{file=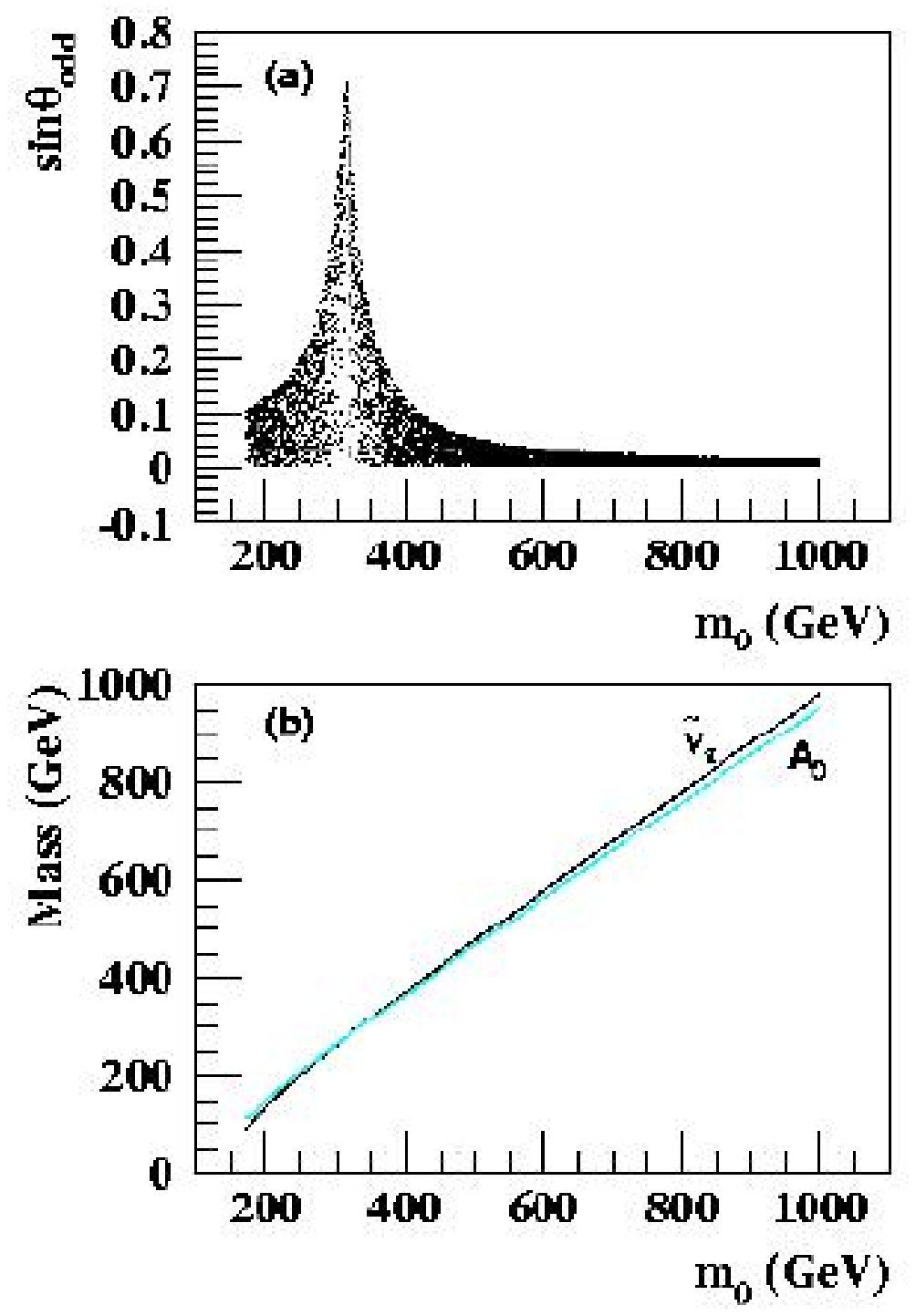,height=11.5cm}}}
\vskip -0.5cm
\caption{
        (a) Mixing of the CP--odd Higgs and the sneutrino and (b) the
        CP--odd Higgs and sneutrino masses as a function of $m_0$ for
        $m_{3/2}=32$ TeV, $\mu<0$ and $\tan \beta = 15$.  }
\label{odd_mo}
\end{figure}

In Fig.~\ref{odd_mo}a we plot tau--sneutrino mixing with the
CP--odd neutral Higgs as a function of $m_0$ for the parameters
indicated above. We find quite large mixings for $m_0\approx 320$ GeV.
In Fig.~\ref{odd_mo}b we show the CP--odd Higgs and tau--sneutrino
masses, which depend almost linearly on $m_0$. Moreover, the value of
$m_0$ at which these two particles have the same mass coincides with
the point of maximum mixing.

\begin{figure}[thb]
\centerline{\protect\hbox{\psfig{file=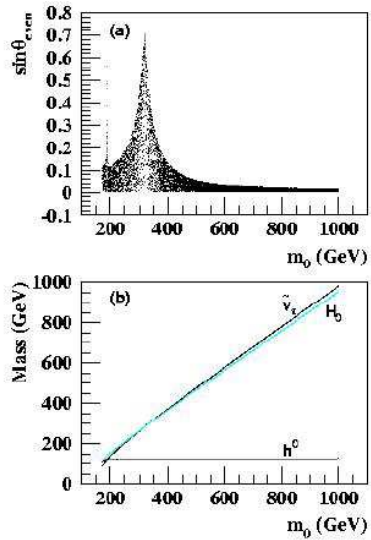,height=12.5cm}}}
\vskip -1.5cm
\caption{
        (a) Mixing between the CP--even Higgs and sneutrino and (b) the
        light and heavy CP--even Higgs masses as well as the sneutrino
        one as a function of $m_0$ for $m_{3/2}=32$ TeV, $\mu<0$ and
        $\tan \beta = 15$.}
\label{even_mo}
\end{figure}

The CP--even tau--sneutrino mixing with the CP--even Higgs is
presented in Fig.~\ref{even_mo}a as a function of $m_0$.  There are
two peaks of high mixing; the main one at $m_0\approx 320$ GeV and a
narrow one at $m_0\approx 180$ GeV. These two peaks have a different
origin, as indicated by Fig.~\ref{even_mo}b, where we plot the masses
of the two CP--even neutral Higgs bosons, $m_h$ and $m_H$, and the
mass of the CP--even tau--sneutrino $m_{\tilde\nu_{\tau}^{even}}$, as a
function of $m_0$. We observe that the broad peak is due to a
degeneracy between the tau--sneutrino and the heavy neutral Higgs
boson and the narrow peak comes from a degeneracy between the
tau--sneutrino and the light neutral Higgs boson. As expected, the
$H^0$ and $\tilde\nu_{\tau}^{even}$ masses grow linearly with $m_0$,
contrary to the $h^0$ mass which remains almost constant.

\begin{figure}
\centerline{\protect\hbox{\psfig{file=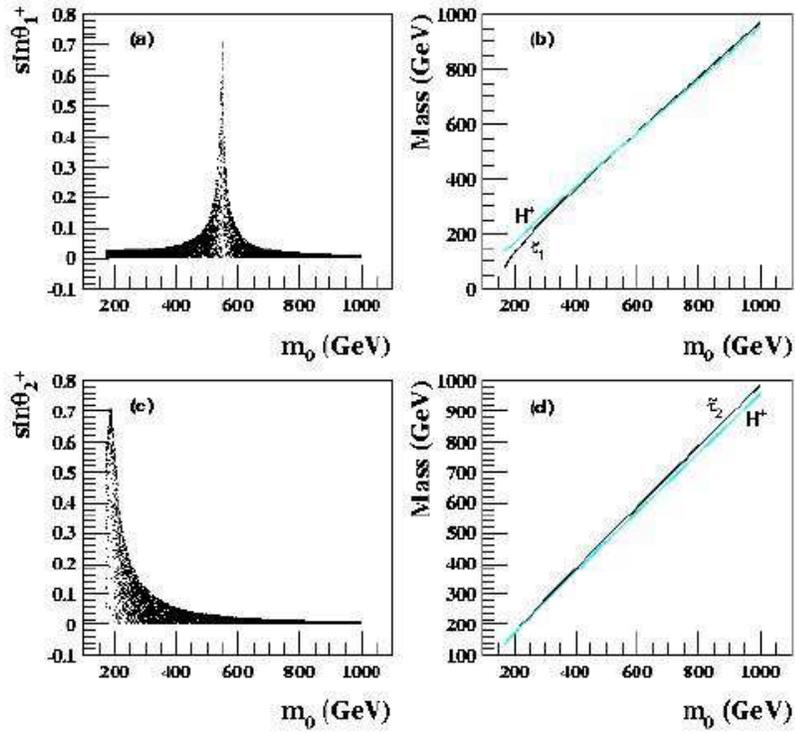,height=17cm}}}
\vskip -1.5cm
\caption{
        (a) Mixing of the charged Higgs with the light stau, (b)
        charged Higgs and light stau masses, (c) mixing of the charged
        Higgs with the heavy stau, and (d) charged Higgs and heavy
        stau masses as a function of $m_0$ for $m_{3/2}=32$ TeV,
        $\mu<0$ and $\tan \beta = 15$.}
\label{charged_mo}
\end{figure}
In Figure \ref{charged_mo}a we display the light stau mixing with the
charged Higgs as a function of $m_0$.  The maximum mixing, obtained at
$m_0\approx 550$ GeV, is the result of a mass degeneracy between the
charged Higgs boson and the light stau. This can be observed in
Fig.~\ref{charged_mo}b where we plot the charged Higgs mass
$m_{H^{\pm}}$ and the light stau mass $m_{\tilde\tau_1}$ as a function
of $m_0$.

In a similar way, we show the heavy stau mixing with charged Higgs as a
function of $m_0$ in Fig.~\ref{charged_mo}c, where we observe a
maximum for the mixing at $m_0\approx 200$ GeV. This large mixing is
due to a degeneracy between the charged Higgs boson and the heavy stau
masses, as can be seen in Fig.~\ref{charged_mo}d. One can notice that
all charged scalars show an almost linear dependency of their mass on
the mass parameter $m_0$.

\begin{figure}
\centerline{\protect\hbox{\psfig{file=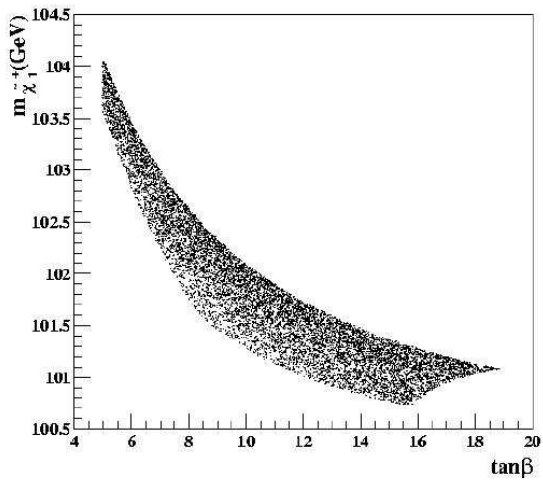,height=11cm}}}
\caption{
 Light chargino mass as a function of $\tan \beta$ for
 $m_{3/2}=32$ TeV, $\mu<0$ and $100<m_0<300$ GeV.  }
\label{chneu_tgb}
\end{figure}

As opposed to the scalar sector, where mixing between the Higgs bosons
and sleptons can be maximum, in the chargino and neutralino sectors
the mixings with leptons are controlled by the neutrino mass being very
small.  Despite this fact, the mixing in the neutralino sector is
sufficient to generate adequate masses for the neutrinos and give rise
to the neutralino decays mentioned in the introduction. Therefore,
in the chargino sector the BRpV--AMSB phenomenology changes very little
with respect to the $R_P$ conserving AMSB. One of the distinctive
features of AMSB that differentiates it from other scenarios of
supersymmetry breaking in the chargino--neutralino sector is the near
degeneracy between the lightest chargino and the lightest
neutralino. This feature remains in BRpV--AMSB as was anticipated in
Fig.~\ref{masses_32000_5}. 
For $m_{3/2}=32$ TeV, $\mu<0$, and $100<m_0<300$
GeV, we show in Fig.~\ref{chneu_tgb} the lightest chargino mass as a
function of $\tan\beta$. The lightest chargino mass has a small
dependence on $\tan\beta$ since its value varies only between 100 and
104 GeV. As in $R_P$ conserving AMSB, the mass difference
$m_{\tilde\chi^+_1}-m_{\tilde\chi^0_1}$ remains small.

\section{Conclusions}
\label{conclusion}

We have shown in the previous sections that our model exhibiting Anomaly
Mediated Supersymmetry Breaking and Bilinear $R_P$ Violation is
phenomenologically viable. In particular, the inclusion of BRpV generates
neutrino masses and mixings in a natural way.  Moreover, the $R_P$ breaking
terms give rise to mixing between the Higgs bosons and the sleptons, which can
be rather large despite the smallness of the parameters needed to generate
realistic neutrino masses. These large mixings occur in regions of the
parameter space where two states are nearly degenerate. Our model also alters
substantially the mass splitting between the scalar taus in a large range of
$\tan \beta$.

The $R_P$ violating interactions render the LSP unstable since it can decay
via its mixing with the SM particles (leptons or scalars). Therefore, the
constraints on the LSP are relaxed and forbidden regions of parameter space
become allowed, where scalar particles like staus or sneutrinos are the LSP.
Furthermore, the large mixing between Higgs bosons and sleptons has the
potential to change the decays of these particles.  These facts have a
profound impact in the phenomenology of the model, changing drastically the
signals at colliders \cite{future}.


\vskip1cm
\noindent{\bf Acknowledgements:}
This work was partially supported by a joint grant from Fundaci\'on Andes and
Fundaci\'on Vitae, by DIPUC, by CONICYT grant No.~1010974, by Funda\c{c}\~{a}o
de Amparo \`a Pesquisa do Estado de S\~ao Paulo (FAPESP), by Conselho Nacional
de Desenvolvimento Cient\'{\i}fico e Tec\-no\-l\'o\-gi\-co (CNPq) 
and by Programa de Apoio a N\'ucleos de Excel\^encia (PRONEX).

\vskip1cm
\appendix
\section{AMSB Boundary Conditions}\label{appa}

The AMSB boundary conditions at the GUT scale for the gaugino masses are
proportional to their beta functions, resulting in
\begin{eqnarray}
M_1&=&{{33}\over 5}{{g_1^2}\over{16\pi^2}}m_{3/2} \; ,
\\
M_2&=&{{g_2^2}\over{16\pi^2}}m_{3/2} \; ,
\\
M_3&=&-3{{g_3^2}\over{16\pi^2}}m_{3/2} \; ,
\end{eqnarray}
while the  third generation scalar masses are given by
\begin{eqnarray}
m_{U}^2 &=& \left(-{88\over 25}g_1^4+8g_3^4+2f_t\hat{\beta}_{f_t}\right)
{m_{3/2}^2\over (16\pi^2)^2}+m_0^2\;,\\
m_{D}^2 &=& \left(-{22\over 25}g_1^4+8g_3^4+2f_b\hat{\beta}_{f_b}\right)
{m_{3/2}^2\over (16\pi^2)^2}+m_0^2\;,\\
m_{Q}^2 &=& \left(-{11\over 50}g_1^4-{3\over 2}g_2^4+
8g_3^4+f_t\hat{\beta}_{f_t}+f_b\hat{\beta}_{f_b}\right)
{m_{3/2}^2\over (16\pi^2)^2}+m_0^2\;,\\
m_{L}^2 &=& \left(-{99\over 50}g_1^4-{3\over 2}g_2^4+
f_\tau\hat{\beta}_{f_\tau}\right) {m_{3/2}^2\over (16\pi^2)^2}+m_0^2\;,\\
m_{E}^2 &=& \left(-{198\over 25}g_1^4+
2f_\tau\hat{\beta}_{f_\tau}\right) {m_{3/2}^2\over (16\pi^2)^2}+m_0^2\;,\\
m_{H_u}^2 &=& \left(-{99\over 50}g_1^4-{3\over 2}g_2^4+
3f_t\hat{\beta}_{f_t}\right) {m_{3/2}^2\over (16\pi^2)^2}+m_0^2\;,\\
m_{H_d}^2 &=& \left(-{99\over 50}g_1^4-{3\over 2}g_2^4+
3f_b\hat{\beta}_{f_b}+f_\tau\hat{\beta}_{f_\tau}\right) 
{m_{3/2}^2\over (16\pi^2)^2}+m_0^2 \;.
\end{eqnarray}
Finally, the $A$--parameters are given by
\begin{eqnarray}
A_t&=&{\hat{\beta}_{f_t}\over f_t}{m_{3/2}\over 16\pi^2}\;,\\
A_b&=&{\hat{\beta}_{f_b}\over f_b}{m_{3/2}\over 16\pi^2}\;,\\
A_\tau &=&{\hat{\beta}_{f_\tau}\over f_\tau}{m_{3/2}\over 16\pi^2}\;,
\end{eqnarray}
where we have defined
\begin{eqnarray}
\hat{\beta}_{f_t} &=& 16\pi^2\beta_t=f_t\left( -{13\over 15}g_1^2-3g_2^2
-{16\over 3}g_3^2+6f_t^2+f_b^2\right)\;,\\
\hat{\beta}_{f_b} &=& 16\pi^2\beta_b=f_b\left( -{7\over 15}g_1^2-3g_2^2
-{16\over 3}g_3^2+f_t^2+6f_b^2+f_\tau^2\right)\;,\\
\hat{\beta}_{f_\tau} &=& 16\pi^2\beta_\tau=f_\tau
\left( -{9\over 5}g_1^2-3g_2^2+3f_b^2+4f_\tau^2\right)\;.
\end{eqnarray}

\section{The Renormalization Group Equations}\label{appb}


Here we present the one--loop renormalization group equations for our model,
assuming the bilinear $R_P$ breaking terms are restricted only to the
third generation.  First, we display the equations for the Yukawa couplings of
the trilinear terms
\begin{equation}
16 \pi^2 \frac{d h_U}{d t} =
h_U \left( 6 h_U^2 + h_D^2 - \frac{16}{3} g_3^2 - 3 g_2^2
- \frac{13}{9} g_1^2 \right)\;,
\end{equation}
\begin{equation}
16 \pi^2 \frac{d h_D}{d t} =
h_D \left( 6 h_D^2 +  h_U^2+  h_{\tau}^2
- \frac{16}{3} g_3^2 - 3 g_2^2
- \frac{7}{9} g_1^2 \right)\;,
\end{equation} 

\begin{equation}
16 \pi^2 \frac{d h_{\tau}}{d t} =
h_{\tau} \left( 4 h_{\tau}^2 + 3 h_D^2
- 3 g_2^2 - 3 g_1^2
\right)\;.
\end{equation}
The corresponding RGE for cubic soft supersymmetry breaking
parameters are given by
\begin{equation}
8 \pi^2 \frac{d A_U}{d t} =
6 h_U^2 A_U + h_D^2 A_D
+ \frac{16}{3} g_3^2 M_3 + 3 g_2^2 M_2
+ \frac{13}{9} g_1^2 M_1\;,
\end{equation}
\begin{equation}
8 \pi^2 \frac{d A_D}{d t} =
 6 h_D^2 A_D +  h_U^2 A_U +  h_{\tau}^2 A_{\tau}
+ \frac{16}{3} g_3^2 M_3+ 3 g_2^2 M_2
+ \frac{7}{9} g_1^2 M_1\;,
\end{equation}
\begin{equation}
8 \pi^2 \frac{d A_{\tau}}{d t} =
 4 h_{\tau}^2 A_{\tau} + 3 h_D^2 A_D
+ 3 g_2^2 M_2 + 3 g_1^2 M_1\;.
\end{equation}
For the soft supersymmetry breaking mass parameters we have
\begin{eqnarray}
8 \pi^2 \frac{d M_{Q}^2}{d t} &=&
h_U^2 ( m_{H_2}^2 + M_Q^2 + M_U^2 + A_U^2)
+ h_D^2 ( m_{H_1}^2 + M_Q^2 + M_D^2 + A_D^2) \cr
&&
- \frac{16}{3} g_3^2 M_3^2 -3 g_2^2 M_2^2 - \frac{1}{9}g_1^2 M_1^2
+ \frac{1}{6}~g_1^2 {\cal S}\;,
\end{eqnarray} 
\begin{equation}
8 \pi^2 \frac{d M_{U}^2}{d t} =
2 h_U^2 ( m_{H_2}^2 + M_Q^2 + M_U^2 + A_U^2)
- \frac{16}{3} g_3^2 M_3^2 - \frac{16}{9}g_1^2 M_1^2
-\frac{2}{3}~g_1^2 {\cal S}\;,
\end{equation}
\begin{equation}
8 \pi^2 \frac{d M_{D}^2}{d t} =
2 h_D^2 ( m_{H_1}^2 + M_Q^2 + M_D^2 + A_D^2)
- \frac{16}{3} g_3^2 M_3^2 - \frac{4}{9}g_1^2 M_1^2
+\frac{1}{3}~g_1^2 {\cal S}\;,
\end{equation}
\begin{equation}
\label{ML}
8 \pi^2 \frac{d M_L^2}{d t} =
h_{\tau}^2 ( m_{H_1}^2 + M_L^2 + M_R^2 + A_{\tau}^2)
-3 g_2^2 M_2^2 - g_1^2 M_1^2
-\frac{1}{2}~g_1^2 {\cal S}\;,
\end{equation}
\begin{equation}
8 \pi^2 \frac{d M_R^2}{d t} =
2 h_{\tau}^2 ( m_{H_1}^2 + M_L^2 + M_R^2 + A_{\tau}^2)
 - 4 g_1^2 M_1^2
+ g_1^2 {\cal S}\;,
\end{equation}
\begin{equation}
8 \pi^2 \frac{d m_{H_2}^2}{d t} =
3 h_U^2 ( m_{H_2}^2 + M_Q^2 + M_U^2 + A_U^2)
- 3 g_2^2 M_2^2 - g_1^2 M_1^2
+ \frac{1}{2}~g_1^2 {\cal S}\;,
\end{equation}
\begin{eqnarray}
\label{MHD}
8 \pi^2 \frac{d m_{H_1}^2}{d t} &=&
3 h_D^2 ( m_{H_1}^2 + M_Q^2 + M_D^2 + A_D^2) +
h_{\tau}^2 ( m_{H_1}^2 + M_L^2 + M_R^2 + A_{\tau}^2)\cr
&&
-3 g_2^2 M_2^2 - g_1^2 M_1^2
- \frac{1}{2}~g_1^2 {\cal S}\;,
\end{eqnarray}
where
\begin{equation}
{\cal S}=  m_{H_2}^2 - m_{H_1}^2 + M_Q^2 -2 M_U^2 + M_D^2
- M_L^2 + M_R^2\;.
\end{equation}
For the bilinear terms in the superpotential we get
\begin{equation}
16 \pi^2 \frac{d \mu}{d t} =
\mu \left(  3 h_U^2 + 3 h_D^2 + h_{\tau}^2
- 3 g_2^2 - g_1^2
\right)\;,
\end{equation}   
\begin{equation}
16 \pi^2 \frac{d \epsilon_3}{d t} =
\epsilon_3 \left(  3 h_U^2 +  h_{\tau}^2
- 3 g_2^2 - g_1^2
\right)\;,
\end{equation}
and for the corresponding soft breaking terms
\begin{equation}
\label{B}
8 \pi^2 \frac{d B}{d t} =
3 h_U^2 A_U + 3 h_D^2 A_D + h_{\tau}^2 A_{\tau}
+ 3 g_2^2 M_2 + g_1^2 M_1\;,
\end{equation}
\begin{equation}
\label{B2}
8 \pi^2 \frac{d B_2}{d t} =
3 h_U^2 A_U + h_{\tau}^2 A_{\tau}
+ 3 g_2^2 M_2 + g_1^2 M_1\;.
\end{equation}
The $g_i$ are the $SU(3)\times SU(2)\times U(1)$ gauge couplings and
the $M_i$ are the corresponding soft breaking gaugino masses.



\end{document}